\documentclass[preprint,prb,aps]{revtex4}
\usepackage{graphicx}
\usepackage{textcomp}
\usepackage{bm}
\begin{document}

\title{Josephson effects in MgB$_{2}$ metal masked ion damage
junctions}
\author{D.-J. Kang\footnote{D.-J. Kang is also with IRC for Nanotechnology, 
University of Cambridge, Cambridge CB3 0HE, UK.}}
\email{djk1003@cam.ac.uk}
\affiliation{Department of Materials Science, University of
Cambridge, Pembroke Street, Cambridge CB2 3QZ, UK}
\author{N. H. Peng}
\author{R. Webb}
\author{C. Jeynes}
\affiliation{Surrey Centre for Research in Ion Beam Applications,
School of Electronics, Computing and Mathematics, University of
Surrey, Guildford, GU2 7XH, UK}
\author{J. H. Yun}
\author{S. H. Moon}
\author{B. Oh}
\affiliation{LG Electronics Institute of Technology, Seoul
137-724, Korea}
\author{G. Burnell}
\author{E. J. Tarte}
\author{D. F. Moore}
\author{M. G. Blamire}
\affiliation{IRC in Superconductivity, University of Cambridge,
Madingley Road, Cambridge CB3 0HE, UK}
\date{\today}
\begin{abstract}
\noindent Ion beam damage combined with nanoscale focused ion beam
direct milling was used to create manufacturable SNS type
Josephson junctions in 100 nm thick MgB$_{2}$ with T$_{C}$ of 38
K. The junctions show non-hysteretic current - voltage
characteristics between 36 and 4.2 K. Experimental evidence for
the dc and ac Josephson effects in MgB$_{2}$ metal masked ion
damage junctions are presented. This technique is particularly
useful for prototyping devices due to its simplicity and
flexibility of fabrication and has a great potential for
high-density integration.
\end{abstract}

\maketitle

\newpage
The recent discovery of superconductivity in MgB$_{2}$ at 39
K\cite{Nagamatsu} is of both fundamental and practical importance
due to the material's attractive properties, including relatively isotropic
superconductivity, large coherence lengths, transparency of
grain boundaries to current flow and the highest transition
temperature ($T_{C})$ found in a simple compound. Both for
applications and basic studies, extensive efforts \cite{Gonnelli,
Brinkman, Zhang, Li, Burnell1, Burnell2} to realize a viable junction
fabrication technology have been made worldwide. Unfortunately,
however, the fabrication of Josephson junctions in thin MgB$_{2}$
films has turned out to be rather difficult because of the
relatively poor fabrication control and lack of suitable barrier
materials for any multilayer type junctions, as well as the
technical challenge of realizing high quality thin films.

Nb-based junctions, with a typical $J_{C}$ of only around 1
kA/cm$^{2}$ have a Stewart-McCumber parameter,$\beta _C $ which is
much larger than 1. As a result, the junctions have a hysteretic
current - voltage ($I$ - $V)$ characteristics, and for
application in dc SQUIDs and RSFQ logic gates they must be
externally shunted to decrease the effective value of $\beta _C $.
Furthermore, maximum clock speed of simple microprocessor that can
be made using such Nb junctions seems to be
limited to less than 25 GHz. \cite{Dorojevets} This number is
barely competive with the current-state-of-art complementary metal
oxide semiconductor and SiGe heterojunction bipolar transistor
technologies for high-performance digital signal processing and
general-purpose computing, even without the burdens of helium
cooling for the operation. However, MgB$_{2}$, has the potential
to overcome these limitations as the material is found to be able
to carry much larger $J_{C, }$ and its
$T_{C}$ is about two times higher than that of the Nb-based
superconductors. This allows one to work safely above 20 or even
30 K, which is comparatively easy with standard cryocoolers.

Ion irradiation has the potential to be used
as a means to modify superconducting properties as well as to
create superconducting weak links.
Bugoslavsky \textit{et al.}\cite{Bugoslavsky} recently showed
that the superconducting properties of MgB$_{2}$ can be strongly
affected by defects and structural disorder created by high energy
ion irradiation. Fabrication of junctions
without interfaces, $i.e.$ weakened structures, by ion or electron
irradiation\cite{Tinchev, Booij} is particularly attractive due to 
its controllability.

In this letter, we report the successful creation of SNS type
MgB$_{2}$ junctions by localized ion implantation. The junctions
display non-hysteretic current - voltage ($I - V)$ characteristics
between 36 and 4.2 K. Microwave-induced steps and an oscillatory
magnetic field dependence of $I_{C}$ were observed. Junctions on
the same chip show nearly identical properties. The junction
parameters were found to be highly controllable through careful
in-situ monitoring of resistance change during ion irradiation and
subsequent rapid thermal annealing (RTA) step.

The films used were 20 nm Au/100 nm MgB$_{2}$ bilayer thin films grown on
(0001) sapphire substrates; details of the
growth has been described elsewhere. \cite{Moon} 20 nm thick Au was 
deposited to protect the MgB$_{2}$ films from degradation during
storage before further processing steps. An additional 430 nm of Au 
was \textit{ex situ} deposited by dc
magnetron sputtering on top of the 20 nm of Au for use as a hard mask during
the ion irradiation step. Tracks 3 $\mu $m wide and contacts were then
patterned by standard optical lithography and broad beam Ar ion milling at
500 V and 10 mA current on a water-cooled rotating stage. The $T_{c}$ of the
tracks was measured before and after photolithography and was found to be
unchanged at around 38 K.

In order to prepare metal mask apertures, the patterned chip was
transferred to the FIB microscope (Philips Electron Optics/FEI
Corporation 200 xP\raisebox{0.06in}{\tiny{\textregistered}} FIB
workstation) with a Ga source. The aperture was defined by writing
a single pixel line cut across the width of
tracks using a 4 pA 30 kV Ga ion beam. The chip was wire-bonded to
enable \textit{in-situ} monitoring of the resistance change during the
FIB milling process to provide us with accurate cutting
depth.\cite{Latif}

After the metal mask apertures were prepared, the sample was
mounted on an ion implanter equipped with a custom cryogenic stage
and shielded to allow ion implantation of the device through a 2
mm aperture. While monitoring the barrier resistance
\textit{in-situ}, the chips were then exposed to a 100 keV ${\rm
H}_2^+$ ion beam with a nominal dose of up to 1 x 10$^{16}$
ions/cm$^{2}$ at 20 K. This temperature was chosen to ensure that
the intended maximum operating temperature of the completed
devices is close to the $T_{c}$ of the bulk after subsequent RTA
processing. RTA was carried out at 300 $^{^{\circ}}$C for 1 min
following a ramp of 60 s. More detailed information on this ion
damage technology has been published elsewhere. \cite{Kang1,
Kang2}

Basic junction characterization was performed between 4.2 K and
$T_{C}$ using a dip probe including magnetic field coils and
microwave antenna. $I - V$ characteristics were obtained in a
quasi-static current-biased measurement. Microwave measurements
were done in the range of 12 - 18 GHz.

Figure 1 shows resistance versus temperature measurements for the same track
before (a) and after ion implantation (b). It is clear from the figure that
FIB processing has no significant effect on $T_{C}$. However, the sharp peak
just before the transition region of the sample was observed after the chip
was irradiated at 20 K. This feature can be ascribed to a current
redistribution along the bilayer voltage leads just before entering the
superconducting phase.

Figure 2 shows the temperature dependence of critical current
($I_{C})$ and normal resistance ($R_{n})$ of the junctions. The
$R_{n}$ was nearly temperature-independent between 36 and 4.2 K
and was about 0.114 $\Omega $ at 4.2 K. This and the temperature dependency
of the $I_{c}$ give evidence for SNS nature of the
junctions. The inset in Fig.~2 is an example of $I - V$
characteristics of the junction at 4.2 K. The $I - V$
characteristics shows resistively shunted junction (RSJ) model
behavior with some excess current. In this case, we have $I_{C}$ =
6.5 mA ($J_{C}$ = 2.17 MA/cm$^{2})$ and $R_{n}$=0.114 $\Omega $
from which a product $I_{C}R_{n}$ = 0.69 mV is obtained. This
value is still small if compared with the value $I_{C}R_{n}=\pi
\Delta $/2 = 3.52$\pi k_{B}T_{C}$/4 = 9.3 mV predicted by the BCS
theory for $T_{C}$ = 39 K. No significant change in the junction
properties was observed over 2 months and a number of thermal
cycles.

Figure 3 shows the magnetic modulation of the $I_{C}$ of a
junction at 34 K. Deviation from the ideal Fraunhofer diffraction
at higher magnetic field in the experimental data implies that
some inhomogeneities exist in local areas of the junction.
However, the data clearly indicates that our barrier shows a good
Josephson junction behavior. The apparently incomplete suppression
of the $I_{C}$ is largely due to the fixed voltage criterion used
to assess the $I_{c}$. We found, however that there is genuinely
an excess critical current for all applied magnetic fields. The
asymmetry and hysteresis of the $I_{C}$ modulation is probably a
consequence of an asymmetrical current distribution due to some
unevenness in the cut made in the mask, or flux trapped during the
long measurement times. The nodes in the critical current occur
every 2 mT. The value of London penetration depth ($\lambda _{L})$
in our irradiated junctions can be obtained from the $I_{C}$(B)
curve by means of the flux density expression $\Delta $B =
\textit{$\Phi $}$_{0}/$(2$\lambda_{L }$+ $l)w$ where \textit{$\Phi
$}$_{0}$ is the flux quantum,$ l$ is the length of junction
barrier, and $w$ is the width of the junction, assuming flux
focusing effects are negligible for a junction of this size. We
found that $\lambda_{L}$ of MgB$_{2}$ was approximately 150 nm.
This is in good agreement with values reported elsewhere.
\cite{Chen, Finnemore, Takano}

To further examine the Josephson effect, 12 GHz microwave were applied to
the junction. Shapiro steps were observed at the expected voltages (V =
\textit{h$\nu $/}2e $\sim $ 25 $\mu $V) as shown in Fig. 4 
despite the low normal resistance of the junction due
to screening by the thick Au mask layer. It is also
likely that such a thick Au mask layer (450 nm) on the top of the junction
barrier will attenuate the interaction between the junction and microwave
radiation. At the frequency ($\sim $ 12 GHz), the skin depth of Au is only
around 680 nm, which is just 1.55 times thicker than the mask.

Improvement in the junction quality following the RTA procedure is
illustrated by comparing an $I_{C}$(B) curve without RTA (Fig. 3
inset) with that after RTA (Fig. 3, main). RTA was taken to
facilitate annealing of most of the low energy defects and
to trap the core defects within a narrow region forming the junction barrier.
Therefore, a reduction in the effective lateral extension of the defect
profile should be expected as well.\cite{Kahlmann} However, even
though we have shown here the successful creation of Josephson
junctions on the MgB$_{2}$, precise mechanism is still unclear.

In conclusion, we have reported the successful creation of
MgB$_{2}$ junctions by localized ion implantation in combination
with focused ion beam direct milling. The junctions show
non-hysteretic $I$ - $V$ characteristics between 36 and 4.2 K.
Clear dc and ac Josephson effects in MgB$_{2}$ metal masked ion
damaged junctions were observed. The product $I_{C}R_{n}$ of our
junctions is small compared to the BCS value but critical
currents up to 6.5 mA at 4.2 K were observed. This technique holds a
promise for prototyping devices due to its simplicity and
flexibility of fabrication and has a great potential for
high-density integration.

This work was supported by the UK Engineering and Physical Sciences Research
Council. H. N. Lee, S. H. Moon and B. Oh are grateful to the Korean Ministry
of Science and Technology under the National Research Laboratory project for
support.

\newpage
\hspace*{-0.35in}\begin{minipage}[t]{2in}
Figure Captions
\end{minipage}
\begin{enumerate}
\item[Fig. 1.] Resistance versus temperature measurements for the same
track of MgB$_{2}$ before (a) and after ion implantation (b).
\item[Fig. 2.] The temperature dependence of $I_{C}$ and $R_{n}$ of
a junction. Inset: $I-V$ characteristics of a junction at 4.2 K.
\item[Fig. 3.] Critical current versus magnetic field for a device
measured at 34 K. Inset shows critical current versus magnetic
field measurement for a device that has not gone through the RTA
treatment.
\item[Fig. 4.] The $I-V$ characteristics of MgB$_{2}$ junction and
Shapiro steps under microwave irradiation of a frequency 12 GHz at
34 K.
\end{enumerate}
\newpage
\begin{figure} [htb]
\includegraphics[width=7.5cm,keepaspectratio=true]{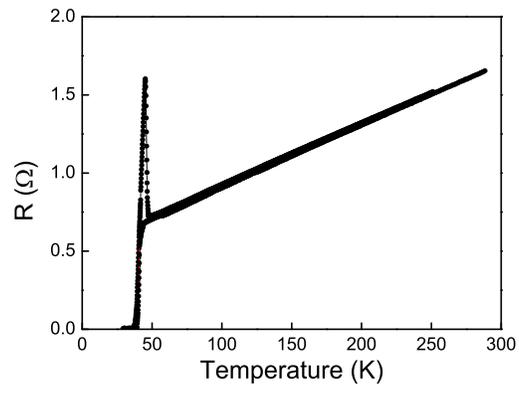}
\vspace{2in}
\caption{\textit{\Large{Kang et al.}}}
\label{fig1}
\end{figure}
\newpage
\begin{figure} [htb]
\includegraphics[width=7.5cm,keepaspectratio=true]{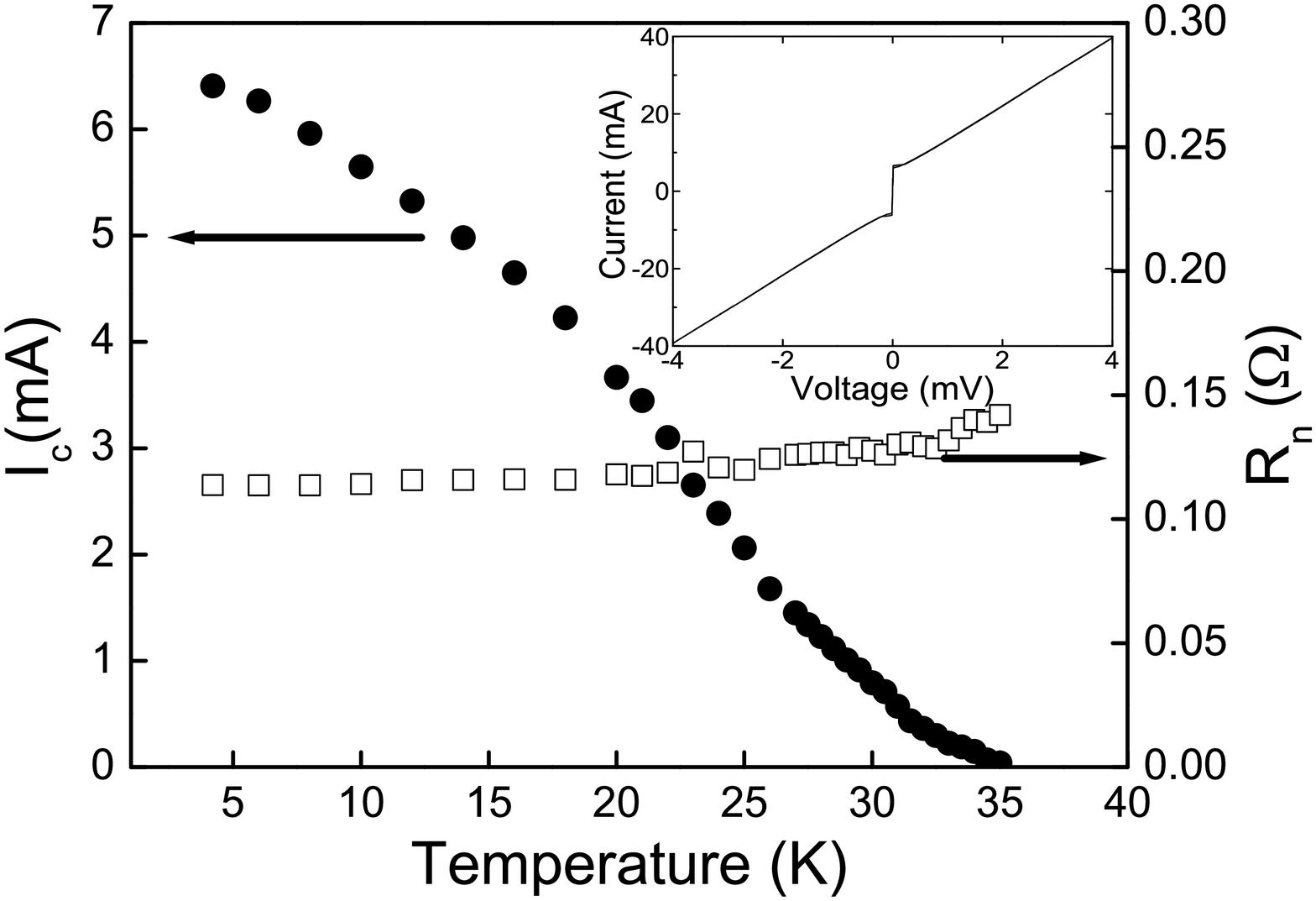}
\vspace{2in}
\caption{\textit{\Large{Kang et al.}}}
\label{fig2}
\end{figure}
\newpage
\begin{figure} [htb]
\includegraphics[width=7.5cm,keepaspectratio=true]{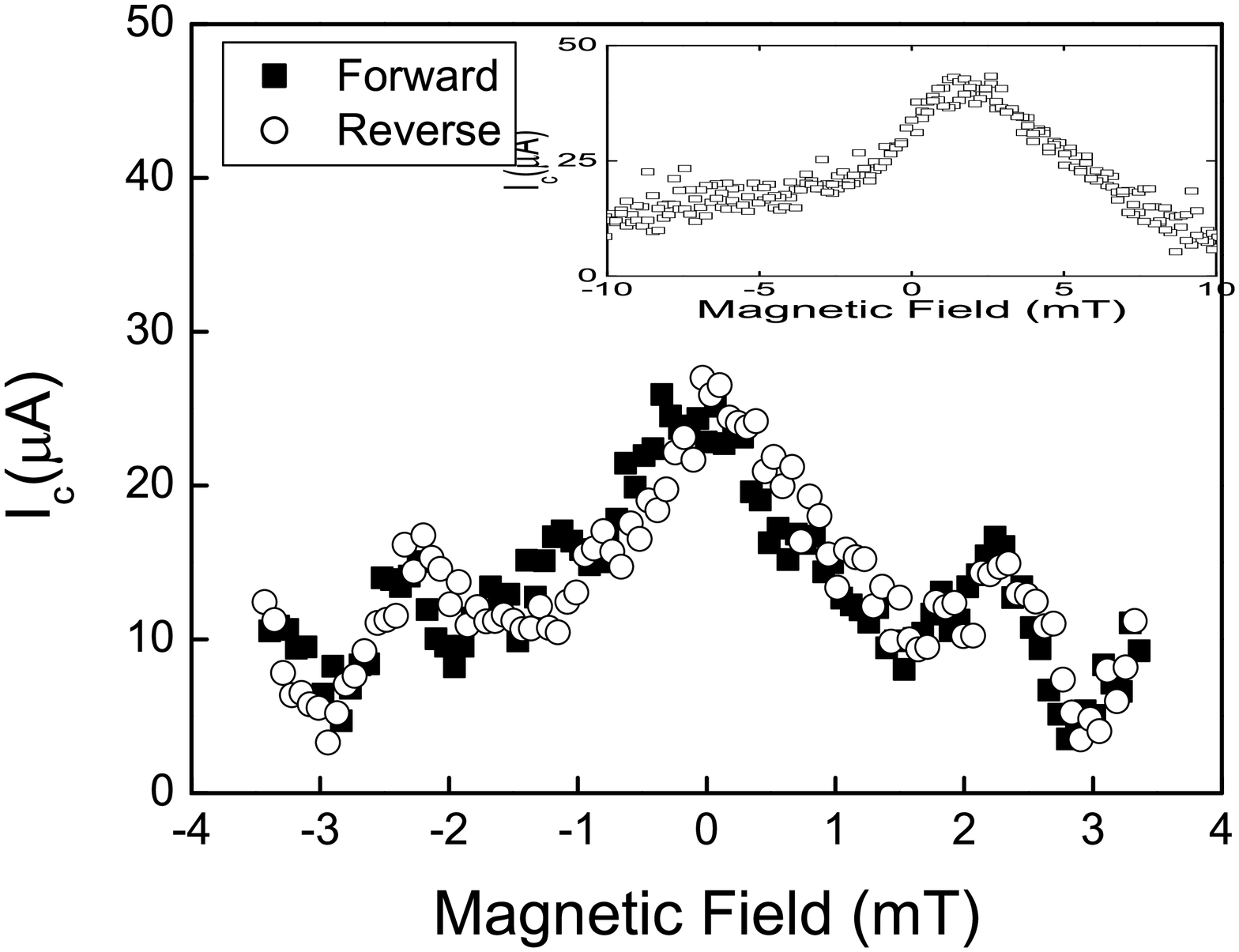}
\vspace{2in}
\caption{\textit{\Large{Kang et al.}}}
\label{fig3}
\end{figure}
\newpage
\begin{figure} [htb]
\includegraphics[width=7.5cm,keepaspectratio=true]{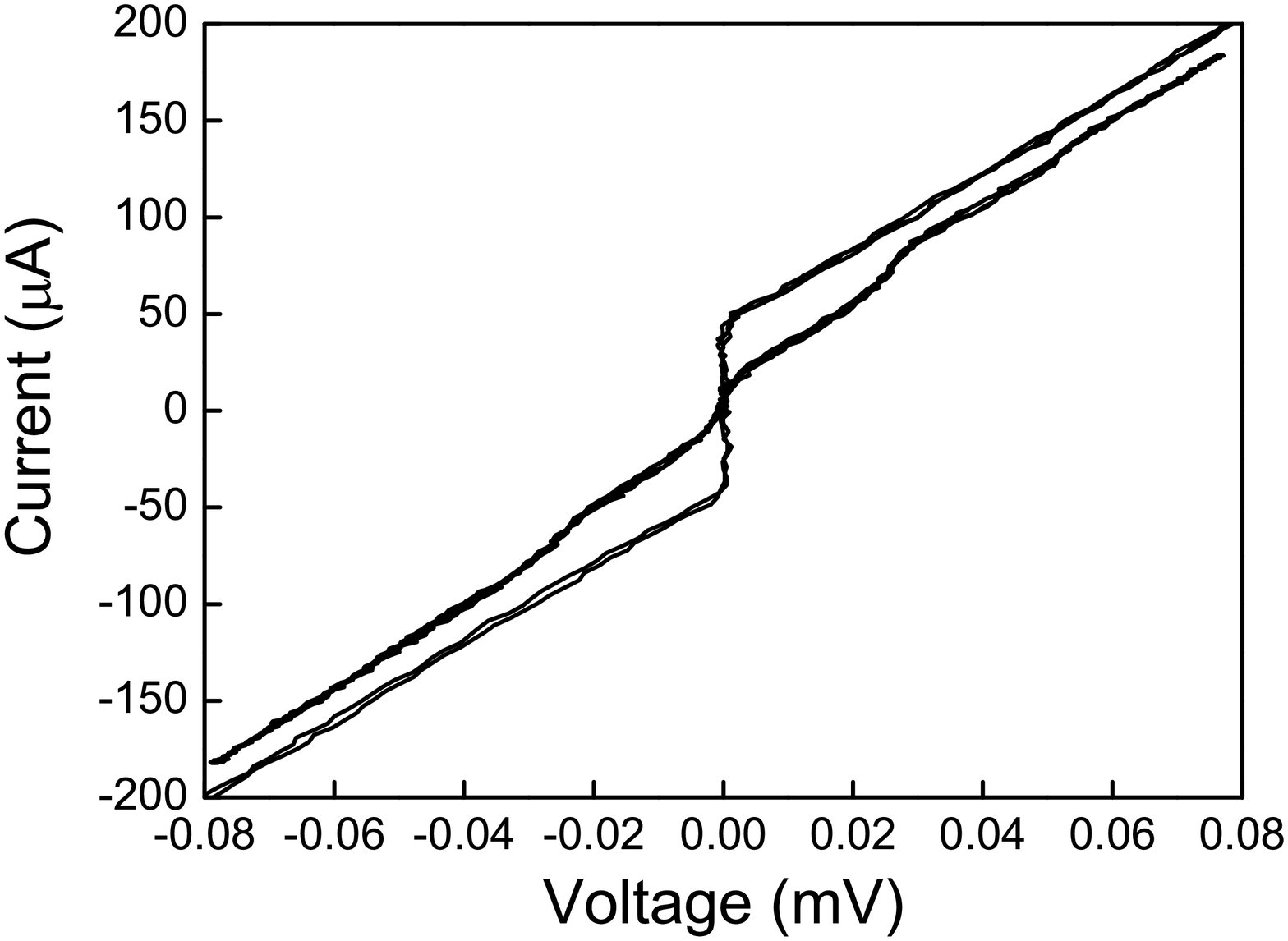}
\vspace{2in}
\caption{\textit{\Large{Kang et al.}}}
\label{fig4}
\end{figure}

\end{document}